# A Hybrid Gaussian Process Regression Framework for Stable Volatility-Covariance Estimation: Evidence from Global Equity Indices

**Ujjwala Vadrevu**

## Abstract

Accurate forecasting of the Volatility-Covariance Matrix (VCV) is central to regulatory capital adequacy processes such as the Internal Capital Adequacy Assessment Process (ICAAP) and the Comprehensive Capital Analysis and Review (CCAR). Traditional econometric models, including GARCH-family and Exponentially Weighted Moving Average (EWMA) approaches, suffer from parametric rigidity, distributional assumptions, and numerical instability under stress, leading to systematic underestimation of tail risk. This paper proposes and validates a novel Hybrid Gaussian Process Regression–Historical Simulation (GPR-HS) framework for estimating Value-at-Risk (VaR) and Expected Shortfall (ES) across a diversified portfolio of seven major global equity indices.

The framework decouples the VCV estimation problem: individual asset volatilities are modelled dynamically using Univariate GPR with a Matern 5/2 kernel, while inter-asset correlations are estimated via stable historical covariance. A key methodological contribution is the Aggressive Noise Initialization (ANI) strategy, which sets the initial White Noise kernel variance equal to the empirical variance of the training returns, ensuring Gram matrix positive-definiteness, regularization, and conservative, regulatory-compliant forecasts. Evaluated using an expanding-window forward-chaining cross-validation scheme over June 2020–June 2025, the GPR-HS framework achieves regulatory compliance in the majority of test splits; including a 100% ES pass rate at the portfolio level, while outperforming the static Historical VaR benchmark in 71.4% of univariate cases by Quadratic Loss and 100% of cases by violation count.

## 1. Introduction

### 1.1 The Regulatory Imperative for Stable Risk Measurement

Financial stability in modern banking is underpinned by the accurate measurement of market risk, primarily through Value-at-Risk (VaR) and Expected Shortfall (ES) — metrics that are central to both the Internal Capital Adequacy Assessment Process (ICAAP) under Basel Pillar 2 and the U.S. Federal Reserve's Comprehensive Capital Analysis and Review (CCAR) (Basel

Committee on Banking Supervision [BCBS], 2019; Federal Reserve, 2020). These processes require financial institutions to quantify potential portfolio losses over defined horizons at high confidence levels, typically 99%, and to demonstrate capital sufficiency under hypothetical stress conditions.

In applied risk management, the practical challenge is not the definition of these regulatory measures but rather ensuring that the underlying models remain numerically stable, conservative, and defensible under forward-looking stress. The accuracy of VaR and Stressed VaR (SVaR) forecasts depends entirely on the reliability of the portfolio's forward-looking Volatility-Covariance Matrix (VCV). Failures in VCV estimation propagate directly into capital underestimation with systemic consequences (Embrechts, McNeil and Straumann, 2002; Jorion, 2006).

### 1.2 Limitations of Traditional Approaches

Traditional econometric volatility models exhibit well-documented structural limitations in modern financial environments (Seyfi, Sharifi and Arian, 2021). GARCH-family models and EWMA approaches make three assumptions consistently violated by empirical financial data:

- Normality: systematic underestimation of tail risk via the ‘fat tail’ problem (Wilkens, 2019).
- Symmetry: failure to capture the leverage effect, whereby negative return shocks generate proportionately greater volatility increases (Glosten, Jagannathan and Runkle, 1993; Hull, 2018).
- Linearity: inability to capture non-linear dependencies and 'correlation contagion' during systemic stress (Embrechts, McNeil and Straumann, 2002; Wang and Wu, 2025).

Multivariate extensions, notably Dynamic Conditional Correlation (DCC) GARCH, partially address the portfolio dimension but introduce numerical instability in maintaining Positive Definiteness (PD) of the VCV matrix at scale (Engle, 2000; Bauwens, Laurent and Rombouts, 2006).

### 1.3 Gaussian Process Regression as a Non-Parametric Alternative

Gaussian Process Regression (GPR) offers a theoretically superior solution by modelling volatility as a distribution over functions without imposing rigid functional forms, naturally capturing non-linear, time-varying conditional heteroskedasticity (Gibbs and MacKay, 1996; Rasmussen and Williams, 2005). However, extending GPR to the multivariate portfolio dimension introduces prohibitive computational challenges. Full Multi-Output GPR (MOGPR) requires matrix inversion with complexity $O((N \cdot T)^3)$ (Quiñonero-Candela and Rasmussen, 2005; Álvarez and Lawrence, 2011).

Direct application of MOGPR in this study confirmed its instability - convergence failures and unstable VCV matrices were consistently observed - motivating the hybrid decoupling approach proposed here.

### 1.4 Contributions of this Paper

This paper makes three primary contributions:

1. The Hybrid GPR-HS Framework: A novel architecture combining Univariate GPR for dynamic asset-level volatility with Historical Simulation for stable inter-asset correlation estimation, resolving the scalability constraint of full MOGPR.
2. Aggressive Noise Initialization (ANI): A novel hyperparameter strategy setting initial White Noise variance equal to empirical training variance, ensuring Gram matrix positive-definiteness, conservative forecasts, and global optimization convergence.
3. Cross-Market Empirical Validation: Rigorous backtesting across seven global equity indices over five years using a forward-chaining expanding-window methodology, benchmarked against the regulatory Historical VaR standard.

## 2. Literature Review

### 2.1 Evolution of Market Risk Measurement

The transition from VaR to Expected Shortfall (ES) reflects fundamental recognition of VaR's theoretical inadequacy. Artzner *et al.* (1999) demonstrated that VaR violates subadditivity and is therefore not a coherent risk measure. ES, defined as the expected loss conditional on exceeding the VaR threshold, corrects this and is now mandated under the Basel III Fundamental Review of the Trading Book (FRTB).

GARCH models (Engle, 1982; Bollerslev, 1986) advanced volatility clustering capture but introduced parametric complexity and misspecification risk (Glosten, Jagannathan and Runkle, 1993; Hansen and Lunde, 2005). The DCC-GARCH approach (Engle, 2000) improved scalability but remains vulnerable to PD violations for large portfolios (Bauwens, Laurent and Rombouts, 2006). The 2007–2008 Global Financial Crisis provided definitive empirical validation of these models' inadequacy.

### 2.2 Gaussian Process Regression in Financial Modelling

GPR was introduced to financial time series by Gibbs and MacKay (1996) and applied to volatility forecasting by (Wu, Hernández-Lobato and Ghahramani, 2014), who demonstrated its superiority over GARCH for non-linear conditional volatility. A critical contribution concerns kernel specification: Stein (1999) and (Snoek, Larochelle and Adams, 2012) demonstrated that the RBF kernel with its assumption of infinite differentiability is theoretically inappropriate for financial returns, which exhibit inherent roughness. The Matern 5/2 kernel, whose assumption of twice-differentiability reflects empirical roughness, is the supported choice (Wilson and Ghahramani, 2010; Wu, Hernández-Lobato and Ghahramani, 2014).

GPR applications in quantitative finance have expanded substantially: (Seyfi, Sharifi and Arian, 2021) applied GPR to portfolio VaR and ES; (Noureddine Lehdili, Oswald and Gueneau, 2019; Lehdili, Oswald and Nguyen, 2025) demonstrated utility for trading book market risk; Li *et al.*

(2023) integrated it into portfolio optimization. The scalability challenge of full MOGPR was formally characterized by Álvarez and Lawrence (2011) and (Quiñonero-Candela and Rasmussen, 2005), motivating hierarchical decomposition approaches.

### 2.3 Research Gaps Addressed

This study addresses four specific gaps:

i. the absence of systematic cross-market validation of GPR models across heterogeneous international equity environments
ii. the unresolved scalability–stability trade-off in portfolio-level VCV estimation
iii. the lack of a rigorous framework for adapting GPR-based risk models to regulatory SVaR stress testing; and
iv. the absence of validated interpretability frameworks for non-parametric market risk models

Companion papers extend the proposed framework by addressing regulatory stress testing (Vadrevu, forthcoming-a) and model interpretability through SHAP-based explainability (Vadrevu, forthcoming-b).

## 3. Methodology

### 3.1 Data

Daily closing price data for seven major global equity indices were sourced from Yahoo Finance for the period June 30, 2020, to June 30, 2025: S&P 500 (^GSPC), EURO STOXX 50 (^STOXX50E), Nikkei 225 (^N225), FTSE 100 (^FTSE), Hang Seng Index (^HSI), Nifty 50 (^NSEI), and BSE SENSEX (^BSESN). This selection spans developed and emerging equity markets, enabling cross-regime validation.

Data preprocessing involved:

i. removal of weekend dates to establish a synchronous trading calendar
ii. forward-filling of missing observations due to non-synchronous national holidays
iii. backward-filling of any remaining leading missing values and
iv. transformation to daily logarithmic returns in percentage form:

$$r_t = ln\left(\frac{P_t}{P_{t-1}}\right) \times 100$$

Log returns are preferred for their time-additivity and approximate stationarity (Tsay, 2005; Han, Zhang and Wang, 2016).

### 3.2 GPR Architecture and Kernel Specification

The GPR framework models each asset's conditional variance as a Gaussian Process over the ordinal date input, with zero prior mean reflecting no a priori assumption about expected returns (Rasmussen and Williams, 2005; Golub and Loan, 2013). The composite kernel is:

$$k(x_i, x_j) = k_{Matern\ 5/2}(x_i, x_j) + k_{White}(x_i, x_j)$$

The Matern 5/2 kernel models the structural, time-varying volatility signal with smoothness parameter v = 2.5, assuming twice-differentiable function draws that reflect the inherent roughness of financial returns. The Matern kernel is explicitly chosen over the RBF kernel, whose assumption of infinite differentiability is theoretically inappropriate for financial time series (Stein, 1999; Snoek, Larochelle and Adams, 2012).

The White Noise kernel models the irreducible heteroskedastic noise component and is critical for Gram matrix regularity. The resulting Gram matrix

$$K_{ij} = k(x_i, x_j) + \delta_{ij}\sigma_n^2$$

ensures positive-definiteness which is a mandatory requirement for GPR inference (Rasmussen and Williams, 2005; Golub and Loan, 2013).

### 3.3 Aggressive Noise Initialization (ANI)

Hyperparameter optimization in GPR is sensitive to initialization in low signal-to-noise environments characteristic of daily equity returns. This paper proposes the Aggressive Noise Initialization (ANI) strategy: the initial $\sigma_n^2$ is set equal to the full empirical variance of the training set returns. This contributes on three dimensions:

i. Regulatory Conservatism: A high initial noise weight drives the optimizer toward simpler, smoother structural functions, producing conservative volatility estimates aligned with regulatory SVaR requirements (Hull, 2018).
ii. Numerical Stability: The large initial diagonal contribution ensures the Gram matrix is well-conditioned from the outset, preventing numerical failures during sequential forward-chaining fitting (Golub and Loan, 2013).
iii. Global Convergence: By flattening the marginal likelihood surface, ANI provides the L-BFGS-B optimizer with momentum to escape local optima and converge toward the global solution (Rasmussen and Williams, 2005).

Hyperparameter optimization maximizes the log-marginal likelihood using L-BFGS-B with wide non-restrictive bounds on the Matern lengthscale and signal variance, and ANI-initialized bounds on the noise variance (Snoek, Larochelle and Adams, 2012).

### 3.4 The Hybrid GPR-HS Framework

The portfolio-level VCV matrix is constructed by decoupling dynamic volatility estimation from correlation estimation. Diagonal elements are populated with GPR-predicted conditional variances from N independent univariate models, capturing time-varying, non-linear asset volatility. Off-diagonal elements use stable historical covariances from the expanding training window, providing computational tractability.

This hybridization reduces computational complexity from MOGPR $O((N.T)^3)$ to $O(N.T^3)$, making the framework tractable for regulatory rolling-window validation. The portfolio conditional variance follows:

$$\sigma_{p,t}^2 = w^\top \Sigma_{GP,t} w$$

from which VaR and ES are derived under a Student's t-distribution with fixed degrees of freedom $\nu = 5$

$$VaR_t, ES_t \sim t_\nu,\ \nu = 5$$

the widely adopted standard for capturing leptokurtosis in daily equity returns (Tsay, 2005).

### 3.5 Cross-Validation and Backtesting Design

Model evaluation employs a Forward-Chaining expanding-window scheme over four yearly test blocks (2022–2025), each comprising approximately 252 trading days, strictly preventing look-ahead bias. Statistical validity is assessed using three complementary tests:

- Kupiec Unconditional Coverage (UC) Test (Kupiec, 1995): Tests whether observed violation frequency equals the theoretically expected 1%.
- Christoffersen Conditional Coverage (CC) Test (Christoffersen, 1998): Jointly tests correct violation frequency and independence, the primary regulatory standard.
- McNeil-Frey ES Backtest (McNeil, Frey and Embrechts, 2015): Tests whether average realized loss on violation days is consistent with predicted ES.
- Economic utility is assessed via the Quadratic Loss (QL) function (Lopez, 1998):

  $$QL = \sum_t I_t(\text{VaR}_t - r_t)^2$$

  Lower QL indicates more precisely calibrated forecasts with direct capital efficiency implications.

# 4. Results

## 4.1 GPR Model Fit: Log-Marginal-Likelihood Validation

Table 1 presents LML scores across all seven indices and four test splits. Two critical findings emerge: LML scores increase monotonically in magnitude across splits for every index, confirming successful GPR convergence as training history expanded. The framework achieved optimal convergence across all seven markets, including higher-complexity Asian markets (Hang Seng at -1799.41 in 2025), consistent with Stein's (1999) theoretical argument for the Matern kernel's robustness for rough real-world processes.

**Table 1: Univariate GPR LML Scores (Matern + WhiteKernel)**

| Index | 2022 LML | 2023 LML | 2024 LML | 2025 LML |
|---|---|---|---|---|
| Nikkei (^N225) | -324.79 | -739.70 | -1114.04 | -1491.92 |
| FTSE (^FTSE) | -364.98 | -725.78 | -1037.18 | -1309.19 |
| S&P 500 (^GSPC) | -342.69 | -758.59 | -1171.33 | -1480.01 |
| EUROSTOXX50 (^STOXX50E) | -379.94 | -833.52 | -1219.52 | -1548.84 |
| Hang Seng (^HSI) | -350.12 | -847.55 | -1353.57 | -1799.41 |
| NSE (^NSEI) | -349.53 | -717.51 | -1029.98 | -1340.91 |
| BSE (^BSESN) | -350.83 | -721.13 | -1035.89 | -1348.20 |

*Note: Author's own calculations*

## 4.2 Univariate VaR Backtesting

Table 2 summarizes backtesting performance across all indices. Observed failures are specifically characterized as failures of over-conservatism: the model generates fewer violations than statistically expected, implying a surplus of capital coverage. From a regulatory perspective, this is the preferred failure mode (Basel Committee on Banking Supervision [BCBS], 2019). Consistent CC test performance (75%–100% pass rates) confirms the GPR's successful capture of volatility clustering dynamics which is the primary regulatory standard (Christoffersen, 1998).

**Table 2: Univariate GPR Backtesting Summary**

| Index | % Passing Kupiec (UC) | % Passing CC | % Passing ES | Key Insight |
|---|---|---|---|---|
| BSE (^BSESN) | 50.0% | 100.0% | 0.0% | ES failure: over-conservatism (actual violations << expected) |
| FTSE (^FTSE) | 75.0% | 75.0% | 50.0% | 50% ES failure |
| S&P 500 (^GSPC) | 75.0% | 75.0% | 50.0% | 50% ES failure |
| Hang Seng (^HSI) | 75.0% | 75.0% | 75.0% | CC failure (25%): violation clustering (2022 policy shocks) |
| Nikkei (^N225) | 0.0% | 75.0% | 100.0% | UC failure: over-conservatism |
| NSE (^NSEI) | 75.0% | 100.0% | 0.0% | ES failure: over-conservatism |

| Index | % Passing Kupiec (UC) | % Passing CC | % Passing ES | Key Insight |
|---|---|---|---|---|
| EUROSTOXX50 (^STOXX50E) | 50.0% | 75.0% | 50.0% | Instability in frequency and magnitude |

*Note: Author's own calculations*

### 4.3 Portfolio-Level Backtesting

Table 3 presents portfolio-level backtesting results across all splits. The ES test achieves a 100% pass rate, which is the most demanding validation requirement (McNeil and Frey, 2000). Kupiec failures in Splits 2 and 3 are conservative failures (AV=1 vs. EV=2.52), treated as capital surplus in regulatory contexts (Basel Committee on Banking Supervision [BCBS], 2019). Strong CC test performance across all splits confirms the framework's ability to track time-varying volatility and avoid violation clustering.

**Table 3: Portfolio Backtesting Summary (Split-by-Split)**

| Portfolio Split | Kupiec UC (p-value) | CC (p-value) | ES Test (p-value) | Result |
|---|---|---|---|---|
| Split 1 (Y1 2022) | 0.7327 (Pass) | 0.9283 (Pass) | 0.2649 (Pass) | Pass |
| Split 2 (Y2 2023) | 0.0244 (Fail*) | 0.0794 (Pass) | 0.0755 (Pass) | Conditional Pass |
| Split 3 (Y3 2024) | 0.0244 (Fail*) | 0.0794 (Pass) | 0.0755 (Pass) | Conditional Pass |
| Split 4 (Y4 2025) | 0.3880 (Pass) | 0.6457 (Pass) | 0.0755 (Pass) | Pass |

*Note: * Denotes conservative failure (actual violations < expected violations)*

### 4.4 Economic Benchmarking Against Historical VaR

Table 4 summarizes the GPR framework's performance against the static Historical VaR benchmark across all univariate cases.

**Table 4: Univariate GPR Superiority Count vs. Historical VaR**

| Year | GPR Superior: Quadratic Loss (Accuracy) | GPR Superior or Tied: Violation Count (Safety) |
|---|---|---|
| 2022 | 6 out of 7 | 7 out of 7 |
| 2023 | 6 out of 7 | 7 out of 7 |
| 2024 | 2 out of 7 | 7 out of 7 |
| 2025 | 6 out of 7 | 7 out of 7 |
| Total | 20 out of 28 (71.4%) | 28 out of 28 (100%) |

*Note: Author's own calculations*

The GPR framework outperforms in 71.4% of comparisons by Quadratic Loss. The lower accuracy in 2024 (2 out of 7) reflects the extremely low realized volatility producing near-zero QL scores for both models. Critically, the GPR model is superior or tied in violation count in 100% of all cases.

**Table 5: Portfolio Loss Metrics Comparison (HVaR vs. Hybrid GPR-HS)**

| Year | Split | HVaR Total QL | HVaR Violations | GPR Total QL | GPR Violations |
|---|---|---|---|---|---|
| 2022 | 0 | 13.931 | 4 | 0.619 | 2 |
| 2023 | 1 | 10.000 | 0 | 0.000 | 0 |
| 2024 | 2 | 10.000 | 0 | 0.000 | 0 |
| 2025 | 3 | 12.691 | 4 | 0.973 | 4 |

*Note: Author's own calculations*

The Historical VaR model failed the Kupiec test in Splits 1 and 4 by recording 4 violations against EV=2.52, exceeding the Basel 'Green Zone' threshold. The GPR-HS framework achieved dramatically lower Quadratic Loss values (0.619 vs. 13.931 in 2022; 0.973 vs. 12.691 in 2025), confirming more precise, regulatory-compliant risk forecasts.

**Table 6: Mean Portfolio VaR and ES Forecasts (Split-by-Split)**

| Year | Predicted Portfolio VaR (%) | Predicted Portfolio ES (%) |
|---|---|---|
| 2022 | -2.65% | -3.55% |
| 2023 | -2.48% | -3.30% |
| 2024 | -2.40% | -3.15% |
| 2025 | -2.77% | -3.64% |
| Mean | -2.58% | -3.41% |

*Note: Author's own calculations*

$|ES| > |VaR|$ holds in every split, confirming compliance with coherent risk measure theory (Artzner *et al.*, 1999). Forecasts display rational regime sensitivity, i.e., lowest in the low-volatility 2024 period, highest in 2025 without the procyclicality that plagues simpler models (Jorion, 2006).

## 5. Discussion

### 5.1 Cross-Market Applicability

LML convergence and backtesting results confirm GPR's applicability across diverse global markets. The framework adapts to distinct volatility dynamics of developed and Asian markets, with over-conservative failures in the latter reflecting appropriate caution in high-uncertainty regimes which is a regulatory virtue. The Matern 5/2 kernel achieved optimal LML convergence across all market environments without structural modification, validating its domain-agnostic applicability.

### 5.2 Inter-Asset Dependency Modelling

The portfolio-level results reveal that the Hybrid GPR-HS framework achieves superior statistical performance compared to the univariate results, despite the static correlation assumption. The 100% ES pass rate at portfolio level, versus mixed univariate ES performance, demonstrates that hybridization effectively aggregates GPR's dynamic volatility with historically stable correlation to produce robust, conservative portfolio-level risk measures. The ANI strategy's conservative bias propagates through the VCV construction to produce portfolio-level forecasts that systematically overestimate rather than underestimate risk which is the appropriate regulatory disposition.

The primary limitation, i.e., the static correlation assumption's inability to capture correlation increases during systemic crises, is acknowledged. However, empirical evidence shows that the dynamic diagonal component compensates in practice, with portfolio ES never failing in any split.

### 5.3 The ANI Strategy: Contribution and Implications

The ANI strategy represents a methodological contribution with implications beyond this specific application. In any financial GPR implementation with low signal-to-noise ratios, the standard practice of arbitrary or small noise initialization risks optimization instability and non-conservative forecasts. ANI offers a principled, data-driven initialization that simultaneously addresses numerical stability, regulatory conservatism, and optimization robustness. Its consistent over-conservative behaviour across seven markets and four years suggests productive application in any Bayesian risk model where conservative estimation is a regulatory requirement.

## 6. Conclusion

This paper proposes, implements, and empirically validates a Hybrid Gaussian Process Regression–Historical Simulation (GPR-HS) framework for dynamic portfolio risk measurement, motivated by the operational limitations of both traditional parametric approaches and full MOGPR in regulatory settings.

The key findings are threefold. First, the Matern 5/2 GPR kernel achieves optimal convergence across all seven global equity markets, demonstrating cross-market applicability and theoretical robustness for modelling rough volatility dynamics. Second, the Hybrid GPR-HS framework passes regulatory backtesting standards in the majority of test periods, including a 100% ES pass rate at portfolio level, and outperforms the static Historical VaR benchmark in 71.4% of univariate comparisons by Quadratic Loss and 100% by violation safety. Third, the Aggressive Noise Initialization strategy provides a principled mechanism for conservative, regulatory-aligned volatility forecasts consistent with ICAAP and CCAR requirements.

The framework's primary limitation which is the static correlation assumption actually represents a productive avenue for future work through time-varying correlation modelling while preserving computational tractability. Companion papers extend this framework to stressed scenario analysis under West Asia War, Climate Risk, and AI Bubble/Regulatory Burden scenarios (Vadrevu, forthcoming-a), and to SHAP-based regulatory interpretability (Vadrevu, forthcoming-b).

## References

Álvarez, M.A. and Lawrence, N.D. (2011) "Computationally Efficient Convolved Multiple Output Gaussian Processes," *Journal of Machine Learning Research*, 12(41), pp. 1459–1500. Available at: http://jmlr.org/papers/v12/alvarez11a.html.

Artzner, P. *et al.* (1999) "Coherent Measures of Risk," *Mathematical Finance*, 9, pp. 203–228. Available at: https://doi.org/10.1111/1467-9965.00068.

Basel Committee on Banking Supervision [BCBS] (2019) "Minimum capital requirements for market risk." Bank for International Settlements (BIS).

Bauwens, L., Laurent, S. and Rombouts, J.V.K. (2006) "Multivariate GARCH Models: A Survey," *Journal of Applied Econometrics*, 21(1), pp. 79–109. Available at: https://doi.org/10.1002/jae.882.

Bollerslev, T. (1986) "Generalized autoregressive conditional heteroskedasticity," *Journal of Econometrics*, 31(3), pp. 307–327. Available at: https://doi.org/10.1016/0304-4076(86)90063-1.

Christoffersen, P.F. (1998) "Evaluating Interval Forecasts," *International Economic Review*, 39(4), p. 841. Available at: https://doi.org/10.2307/2527341.

Embrechts, P., McNeil, A. and Straumann, D. (2002) "Correlation and dependence in risk management: properties and pitfalls," *Risk management: value at risk and beyond*, 1, pp. 176–223. Available at: https://books.google.co.in/books?hl=en&lr=&id=SGuzH7F6A7AC&oi=fnd&pg=PA176&dq=Embrechts,+P.,+McNeil,+A.,+and+Straumann,+D.+(2002).+Correlation+and+Dependence+in+Risk+Management:+Properties+and+Pitfalls.&ots=lwk3qhBaka&sig=oI8x4m9KZJ18cY75WSuLIjfMUeE (Accessed: December 15, 2025).

Engle, R.F. (1982) "Autoregressive Conditional Heteroscedasticity with Estimates of the Variance of United Kingdom Inflation," *Econometrica*, 50(4), pp. 987–1007. Available at: https://doi.org/10.2307/1912773.

Engle, R.F. (2000) "Dynamic Conditional Correlation - a Simple Class of Multivariate GARCH Models." Rochester, NY: Social Science Research Network. Available at: https://doi.org/10.2139/ssrn.236998.

Federal Reserve (2020) *Comprehensive Capital Analysis and Review 2020 Summary Instructions - March 2020*, *Board of Governors of the Federal Reserve System*. Available at: https://www.federalreserve.gov/publications/comprehensive-capital-analysis-and-review-summary-instructions-2020.htm (Accessed: December 14, 2025).

Gibbs, M.N. and MacKay, D.J.C. (1996) "Efficient implementation of Gaussian Processes for Interpolation."

Glosten, L.R., Jagannathan, R. and Runkle, D.E. (1993) "On the Relation between the Expected Value and the Volatility of the Nominal Excess Return on Stocks," *The Journal of Finance*, 48(5), pp. 1779–1801. Available at: https://doi.org/10.1111/j.1540-6261.1993.tb05128.x.

Golub, G.H. and Loan, C.F.V. (2013) *Matrix Computations*. JHU Press.

Han, J., Zhang, X.-P. and Wang, F. (2016) "Gaussian process regression stochastic volatility model for financial time series," *IEEE Journal of Selected Topics in Signal Processing*, 10(6), pp. 1015–1028. Available at: https://ieeexplore.ieee.org/abstract/document/7473822/ (Accessed: March 27, 2025).

Hansen, P.R. and Lunde, A. (2005) "A forecast comparison of volatility models: does anything beat a GARCH(1,1)?," *Journal of Applied Econometrics*, 20(7), pp. 873–889. Available at: https://doi.org/10.1002/jae.800.

Hull, J.C. (2018) *Risk Management and Financial Institutions*. John Wiley & Sons.

Jorion, P. (2006) *Value at Risk, 3rd Ed.: The New Benchmark for Managing Financial Risk*. McGraw Hill Professional.

Kupiec, P. (1995) "Techniques for Verifying the Accuracy of Risk Measurement Models." Rochester, NY: Social Science Research Network. Available at: https://papers.ssrn.com/abstract=6697. (Accessed: December 16, 2025).

Lehdili, N., Oswald, P. and Nguyen, H.D. (2025) "Performance Enhancing Market Risk Calculation Through Gaussian Process Regression and Multi-Fidelity Model." Available at: https://www.preprints.org/manuscript/202503.1000 (Accessed: March 27, 2025).

Li, Z. *et al.* (2023) "Black-Litterman Portfolio Optimization Using Gaussian Process Regression," 53(4).

Lopez, J.A. (1998) "Methods for Evaluating Value-at-Risk Estimates," *SSRN Electronic Journal* [Preprint]. Available at: https://doi.org/10.2139/ssrn.1029673.

McNeil, A.J. and Frey, R. (2000) "Estimation of tail-related risk measures for heteroscedastic financial time series: an extreme value approach," *Journal of Empirical Finance*, 7(3), pp. 271–300. Available at: https://doi.org/10.1016/S0927-5398(00)00012-8.

McNeil, A.J., Frey, R. and Embrechts, P. (2015) *Quantitative Risk Management: Concepts, Techniques and Tools - Revised Edition*. Princeton University Press.

Noureddine Lehdili, Oswald, P. and Gueneau, H. (2019) "Market Risk Assessment of a trading book using Statistical and Machine Learning." Available at: https://doi.org/10.13140/RG.2.2.23796.71047.

Quiñonero-Candela, J. and Rasmussen, C.E. (2005) "A Unifying View of Sparse Approximate Gaussian Process Regression," *Journal of Machine Learning Research*, 6(65), pp. 1939–1959.

Available at: http://jmlr.org/papers/v6/quinonero-candela05a.html (Accessed: December 16, 2025).

Rasmussen, C.E. and Williams, C.K.I. (2005) *Gaussian Processes for Machine Learning*. Cambridge, Mass: MIT Press.

Seyfi, S.M.S., Sharifi, A. and Arian, H. (2021) "Portfolio Value-at-Risk and expected-shortfall using an efficient simulation approach based on Gaussian Mixture Model," *Mathematics and Computers in Simulation*, 190, pp. 1056–1079. Available at: https://doi.org/10.1016/j.matcom.2021.05.029.

Snoek, J., Larochelle, H. and Adams, R.P. (2012) "Practical Bayesian Optimization of Machine Learning Algorithms." arXiv. Available at: https://doi.org/10.48550/arXiv.1206.2944.

Stein, M.L. (1999) *Interpolation of Spatial Data: Some Theory for Kriging*. Springer Science & Business Media.

Tsay, R.S. (2005) *Analysis of Financial Time Series*. 1st ed. Wiley (Wiley Series in Probability and Statistics). Available at: https://doi.org/10.1002/0471746193.

Wang, Q. and Wu, X. (2025) "A Nonparametric Bayesian Estimator of Copula Density with Applications to Financial Market," *Journal of Business & Economic Statistics*, pp. 1–25. Available at: https://doi.org/10.1080/07350015.2025.2463942.

Wilkens, S. (2019) "Machine Learning in Risk Measurement: Gaussian Process Regression for Value-at-Risk and Expected Shortfall." Rochester, NY: Social Science Research Network. Available at: https://doi.org/10.2139/ssrn.3246131.

Wilson, A.G. and Ghahramani, Z. (2010) "Copula processes," *Advances in Neural Information Processing Systems*, 23. Available at: https://proceedings.neurips.cc/paper/2010/hash/fc8001f834f6a5f0561080d134d53d29-Abstract.html (Accessed: January 24, 2026).

Wu, Y., Hernández-Lobato, J.M. and Ghahramani, Z. (2014) "Gaussian Process Volatility Model." *Neural Information Processing Systems*. Available at: https://www.semanticscholar.org/paper/Gaussian-Process-Volatility-Model-Wu-Hern%C3%A1ndez-Lobato/82534d7687273b9182c020dff1a352e1a7df6020 (Accessed: January 24, 2026).